# OpenSIP: Toward Software-Defined SIP Networking

Ahmadreza Montazerolghaem, *Student Member, IEEE,* Mohammad Hossein Yaghmaee, *Senior Member, IEEE,*
Alberto Leon-Garcia, *Fellow, IEEE,*

*Abstract*—VoIP is becoming a low-priced and efficient replacement for PSTN in communication industries. With a widely growing adoption rate, SIP is an application layer signaling protocol, standardized by the IETF, for creating, modifying, and terminating VoIP sessions. Generally speaking, SIP routes a call request to its destination by using SIP proxies. With the increasing use of SIP, traditional configurations pose certain drawbacks, such as ineffective routing, un-optimized management of proxy resources (including CPU and memory), and overload conditions. This paper presents OpenSIP to upgrade the SIP network framework with emerging technologies, such as SDN and NFV. SDN provides for management that decouples the data and control planes along with a software-based centralized control that results in effective routing and resource management. Moreover, NFV assists SDN by virtualizing various network devices and functions. However, current SDN elements limit the inspected fields to layer 2-4 headers, whereas SIP routing information resides in the layer-7 header. A benefit of OpenSIP is that it enforces policies on SIP networking that are agnostic to higher layers with the aid of a Deep Packet Inspection (DPI) engine. Among the benefits of OpenSIP is programmability, cost reduction, unified management, routing, as well as efficient load balancing. The present study implements OpenSIP on a real testbed which includes Open vSwitch and the Floodlight controller. The results show that the proposed architecture has a low overhead and satisfactory performance and, in addition, can take advantage of a flexible scale-out design during application deployment.

## I. INTRODUCTION

SESSION Initiation Protocol (SIP) is extensively deployed for significantly growing session-oriented applications on the Internet, such as Internet telephony or voice over IP [1]. As a signaling protocol for real-time communication, SIP performs user location, session establishment, and session management. With the rapid growth of the Internet and Cloud computing, all phones and other mobile devices will soon have to support SIP as their signaling protocol for multimedia sessions [2]. However, the traditional SIP network architecture renders the management of these networks both complicated and expensive. Hence, dealing with such challenges increases operational expenses (Opex). Moreover, the additional complexities and high modification characteristic of the SIP network create instability in the ecosystem. The above-mentioned factors demonstrate the need for a new approach to network management. The problem is that a large SIP network mainly consists of hardware-centric proxies from various vendors with large scale topologies requiring monitoring and configuration. In addition, with the expansion of SIP networks, fundamental mechanisms, such as call request routing [3], have lost their significance, thus posing difficulties, such as the SIP proxy overload. Therefore, there is a vital need for the softwarization of SIP. To this end, Software-Defined Networking (SDN) [4] is quite handy. With SDN and the aid of software-based controllers alongside pre-designed APIs (e.g. OpenFlow protocol [5]), an entire network and its switches may be controlled and programmed as a unified network.

Ahmadreza Montazerolghaem and Mohammad Hossein Yaghmaee are with the Department of Computer Engineering, Ferdowsi University of Mashhad, Mashhad, Iran (e-mail: ahmadreza.montazerolghaem@alumni.um.ac.ir, hyaghmae@um.ac.ir).

Alberto Leon-Garcia is with the Department of Electrical and Computer Engineering, University of Toronto, Toronto, Canada (e-mail: alberto.leongarcia@utoronto.ca)

The architecture of SDN and the OpenFlow protocol allow the separation of data and the control plane, which results in controlling the network and making it even smarter. On the other hand, there is the possibility of separating the network infrastructure from the applications. The goal of the current study is to separate the main functionalities of the SIP networks, which are the mastermind of the network, from the data infrastructure with its currently excessive complexity. The present work then forms these functionalities into a software logically centralized controller referred to as OpenSIP. This problem is approached in three consecutive steps. Considering that the main body of SIP networks is the proxies, the first approach attempts to achieve load balancing amongst SIP proxies using SDN technology. The second approach takes a step forward and tries to move the SIP proxy functionalities from the data to the control plane. In doing so, the data plane is eradicated from the SIP proxy equipment and the complications involved in management. Finally, the third approach seeks to virtualize the SIP proxies in such a way that SDN manages the virtual proxy infrastructure by employing the Network Function Virtualization (NFV) concept [6]. From softwarization aspects, virtualization maximizes the efficacy of resources. This fact, together with optimization, flexibility, migration of virtual machines, and the possibility of reducing or adding resources, are other benefits of virtualization. The flexibility of the virtual SIP network is related to the input load and may increase accordingly. In fact, resources are easily assigned to various parts of the SIP network.

To our knowledge, OpenSIP is the first work able to practically improve the performance of SIP networks on a softwarization bed by utilizing the concepts of SDN and NFV. There are certainly challenges along the way. The most critical challenge is that current SDN protocols, such as OpenFlow, concentrates on layers 2 to 4 [7], while SIP is a protocol for layer 7. The SIP proxy needs a higher layer to perform its operations, such as routing and registration. This is the reason why Deep Packet Inspection (DPI) technology is employed. DPI is a classification/detection technique capable of detecting SIP messages and providing layer 7 information for decision making.

### A. Contributions

The most important contributions of the current study are summarized as follows:

1) The proposal of three novel approaches for improving SIP network by employing the concepts of SDN and NFV
   - First approach: Partial SDN-based architecture (OpenSIP$^{\text{Partial}}$)
   - Second approach: Full SDN-based architecture (OpenSIP$^{\text{Full}}$)
   - Third approach: Software-defined NFV-based architecture (OpenSIP$^{\text{NFV+}}$)
2) The extension of the OpenFlow protocol, SDN controller, and switches in such a way so as to insure they are constantly aware of higher layer information
3) The implementation and evaluation of the performance of the proposed approaches on a real testbed and under various scenarios

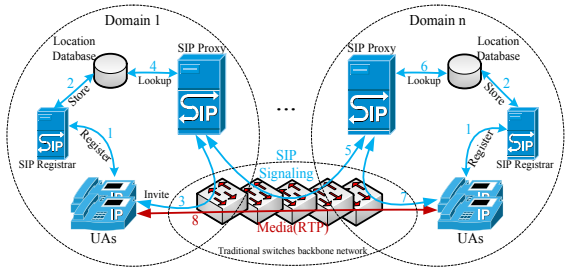

Fig. 1. Registration and call initiation in SIP

### B. Organization

In Section II, a background is presented on SIP networks, their setbacks, and on SDN networks. Section III introduces the three approaches leading to OpenSIP. Section IV is dedicated to assessing these approaches. Section V offers an overview of related works and finally Section VI concludes the study and presents ideas for possible future work.

## II. BACKGROUND

### A. SIP Networking

SIP is an application layer signaling protocol (layer 7 protocol) that is used to initiate, modify, and terminate multimedia sessions. In other words, SIP is a coordinator between SIP endpoints. The SIP network consists of two entities, namely the User Agent (UA) and server. UA is divided into the User Agent Client (UAC or caller) and User Agent Server (UAS or callee) which generate the request and the response messages, respectively. The request messages are produced and transmitted by the UAC and contain information about the sender and receiver. Responses confirm that a request was received and processed and contain the processing status as well. Servers are also divided into registrar and proxy servers. Registrar servers are responsible for registering user information, whereas proxy servers search for the intended user and route the messages [8]. The most important message in this context is the call setup request or `Invite`. This is pivotal as routing for this request and the remaining request and response messages follow the same path as the one created for this message if the proxies decide to include their identities in the *Record-Route* header field. Therefore, the `Invite` message has the highest processing overhead in the session initiation phase. In fact, a SIP proxy receives an `Invite` from a user agent or another proxy and acts on behalf of the user agent in forwarding or responding to the request. Just as a router forwards IP packets at the IP layer, so a SIP proxy forwards SIP messages at the application layer. A proxy server typically has access to a database or a location service to aid it in processing the request (determining the next hop). Databases may contain SIP registrations, presence information, or any other type of information related to the user location [8]. Fig. 1 shows the registration and session initiation steps in a typical SIP configuration with $n$ domains. The session invitation usually traverses a set of proxies until it finds one which knows the actual location of the callee. Such a proxy will forward the session invitation directly to the callee and the callee then accepts or declines the session invitation.

Fig. 2 presents a typical SIP trapezoid topology and standard SIP call signaling which consists of the `Invite`-`Bye` message sequence. A session is initiated when the caller sends an `Invite` request to the callee. The request is then routed through several SIP proxies (multihop). The `100 Trying` response is sent back to the previous hop for confirmation of the request. Once a callee receives the `Invite` request, it returns a `180 Ringing` response to the caller. The path chosen for this action is the same as the path for sending the `Invite` message. Later, when accepting the call, the callee sends back a `200 Ok` message in response. The caller, in turn, acknowledges the receipt of the `200 Ok` by sending an `Ack` to the callee. After this three-way handshake, a Real-time Transport Protocol (RTP), including voice and video, is independently established between the caller and the callee without having been routed through the SIP proxies. Hence, the load on SIP proxies is the signaling messages. The session is then terminated when one party sends a `Bye` request which is followed by a `200 Ok` response from the other party [9].

SIP messages are transported independently of the transport layer network, even though they are typically transported in a UDP datagram. Each message consists of *First-Line*, *Message-header*, and *Message-body* (see Fig. 3) [8]. The first line identifies the type of message (request or response). A SIP request may contain one or more *Via* header fields which record the path of the request. These are later employed to route SIP responses in exactly the same way. The `Invite` message contains exactly one *Via* header field which is created by the UA sending the request. Whether or not the UA is running on host 195.37.77.100 and port 5040 can be determined from the *Via* field. The *From* and *To* header fields identify the caller and callee of the invitation. The *From* header field contains a *tag* parameter which serves as a dialog identifier, in which a dialog defines the peer-to-peer SIP relationship between two UAs. Dialogs facilitate the sequencing and routing of messages between SIP endpoints. The *Call-ID* header field is a dialog identifier that identifies messages from the same call. Moreover, the order of requests is preserved by *CSeq*. Given that it is likely that requests are sent via an unreliable transport, which might re-order messages, attaching a sequence number to the messages is necessary. This allows the recipient to identify retransmissions and requests that are out of order. The *Contact* header field determines the *IP address* and *port number*, on which the sender awaits further caller requests. The detail of other header fields is irrelevant to this discussion and so are not addressed further here. Note that the message header is delimited from the message body by an empty line. In addition the message body of the *Invite* request contains a description of the media type accepted by the caller and is encoded in the Session Description Protocol (SDP) [8].

### B. Problems Associated with SIP

As mentioned earlier, the most critical task of a SIP proxy is to route session invitations closer to the callee. Due to an absence of a SIP network global view, it is possible that any one of the SIP proxies faces an overload when using this type of hop-by-hop routing. This is because a proxy is significantly affected by its processing and

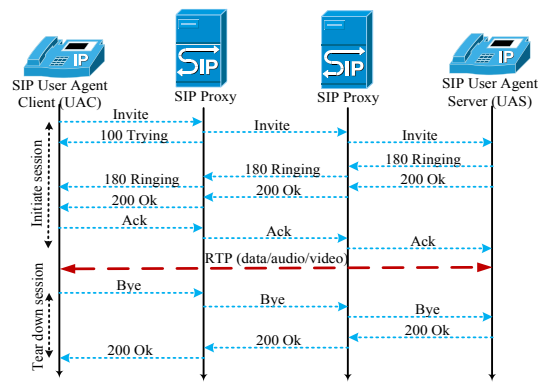

Fig. 2. Basic SIP call flow

```
INVITE sip:717@iptel.org SIP/2.0
Via: SIP/2.0/UDP 195.37.77.100:5040;rport
Max-Forwards: 10
From: "jiri" <sip:jiri@iptel.org>;tag=76ff7a07-c091-4192-84a0-d56e91fe104f
To: <sip:jiri@at.iptel.org>
Call-ID: d10815e0-bf17-4afa-8412-d9130a793d96@213.20.128.35
CSeq: 2 INVITE
Contact: <sip:213.20.128.35:9315>
User-Agent: Windows RTC/1.0
Proxy-Authorization: Digest username="jiri", realm="iptel.org",
  algorithm="MD5", uri="sip:jiri@at.iptel.org",
  response="3cef75390000000177132805ae1b8b7f0d742da1feb5753c",
  nonce="3cef75390000000177132805ae1b8b7f0d742da1feb5753c",
  b03b3e0643bda70f"
Content-Type: application/sdp
Content-Length: 451

v=0
o=jku2 0 0 IN IP4 213.20.128.35
s=session
c=IN IP4 213.20.128.35
b=CT:1000
t=0 0
m=audio 54742 RTP/AVP 97 111 112 6 0 8 4 5 3 101
a=rtpmap:97 red/8000
```

Fig. 3. The structure of a SIP network message

memory resources, which explains why suboptimal routing causes saturation and overload[1]. The overload then results in a delay while establishing calls or may cause a loss of some calls. In order to achieve maximum network capacity and avoid overload, extra care should be given to effectively manage SIP networks. This is possible only if centralized managing and optimal routing are deployed. To this end, a SIP proxy can be developed based on softwarization concept (in SDN) or as a VNF.

### C. Software-Defined Networking

In spite of the widespread application of distributed protocol and control that operate inside network devices (e.g. switches, routers, etc.), the management of a combination of such elements (e.g. SIP networks) is quite challenging given their vertical implementation. Consequently, the control plane, which is responsible for traffic control decisions (e.g. routing), and the data plane, which steers traffic based on such decisions, are both implemented inside the network devices. This reduces flexibility as well as innovation. Moreover, operators must individually configure each one of the devices by using low order commands. SDN is an emerging network concept that intends to remove current limitations by eliminating vertical structures, separating the logical network plane from the network devices, and by spreading centralized network control. Using SDN, the network functionalities can be developed on a software bed regardless of the product brand. Moreover, it is possible to obtain a global view of the network status and achieve high flexibility as well as simple and unified management [11], [12]. Fig. 4 presents a view of SDN architecture. The data plane in this architecture is comprised of network devices, such as forwarding, switching, and others which are deprived of any control or software centers for automatic decision making. The smart element of the network is in the SDN software controllers which retain the overall architecture of the network. The application plane contains a set of applications, such as routing, firewall, load balancing, and so on. The communication protocol between the planes is a series of standard Open APIs, including OpenFlow [4], [5]. Whenever such interfaces exist, the controller is able to dynamically program the inhomogeneous devices of the network. This protocol uses flow tables containing three fields: a *header* or *matching*, which is a series of information from the headers (up to layer 4), an *action* which is performed for the matched packets, and *counters* that preserve the statistics of the matched packets [13]. By varying these fields, the OpenFlow upper layer model realizes various devices (such as routers, firewalls, etc.) in the data plane. For example, an OpenFlow switch has flow tables whose *header* fields feature layer 2 information. As soon as a packet enters this switch, the packet *header* fields are matched with that of the flow table. If matching is verified, the *action* is performed on the packet. The actions consist of 1) sending the packets to a specific output port, 2) encapsulating the packets and sending them to the controller, and 3) dropping the packets, and so on. However, if matching is not verified, the packet is encapsulated in a `Packet-In` message and sent to the controller. When the controller receives the `Packet-In` message, one or more applications running on the controller may process the message and install rules in the flow table via a `Flow-Mod` message. This allows the later packets of the flow to be processed by the switch. Moreover, the controller can inject packets into the data plane of a particular switch. This is possible with the `Packet-Out` message, which carries a packet that is injected into the switch. Unfortunately, the current switches and OpenFlow limit the inspected fields to layer 2-4 headers.

A powerful complement to SDN is NFV [11], [14]. The aim of virtualization is to virtually implement network devices and functions, which are then called Virtualized Network Functions (VNF) [15]. A VNF is typically made up of one or several virtual machines that operate a specific software on servers or even on a cloud infrastructure in Cloud computing in order to present a specific function (rather than a separate software considered for each function). Hence, NFV separates network functions from hardware. This approach offers the possibility of softwarizing the functions while virtual machines control them [6].

## III. OPENSIP

The goal of the current study is to approach software-defined SIP networking in three separate coherent steps. Fig. 5(a) shows a domain of a current SIP network forming a network of traditional switches, a communication infrastructure between UAs, and SIP proxies. Note that the data and control plane are not separate. As the number of UAs increase, it becomes more likely that load balancing occurs between the proxies of those domains. Therefore, as time goes on, some proxies will face overload while others are relatively idle. This situation results in a delay when establishing a call, the loss of a call entirely, or the inefficient use of proxy resources.

The first approach presented by the present work intends to use the current SDN, with minor alterations, for load balancing among SIP proxies (Fig. 5(b)). In the second approach, the control plane of SIP proxy is decoupled from the infrastructure layer and SIP routing is centralized and softwarized (Fig 5(c)). So, the main idea is to move the control plane of SIP proxy into a central controller that is in charge of taking all routing decisions in the SIP network. Finally, a SIP proxy on the data plane is virtually constructed in the third approach. This is done in such a way that the management of the communication infrastructure (switches) of virtual switches is on SDN (Fig. 5(d)). The detail for each approach is given below.

Fig. 4. Overview of SDN

---

[1] [10] proved that the problem of overload control in a SIP network with limited resources is NP-hard.

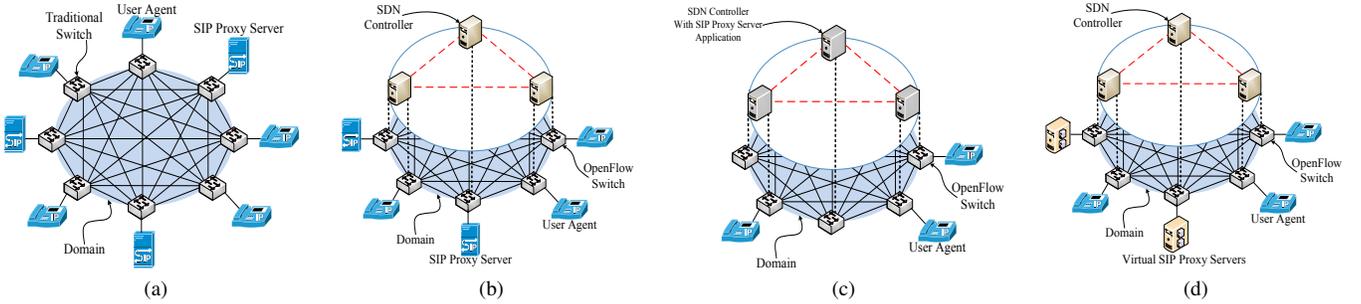

Fig. 5. (a) An overview of a traditional network (b) OpenSIP$^{\text{Partial}}$ (c) OpenSIP$^{\text{Full}}$ (d) OpenSIP$^{\text{NFV+}}$

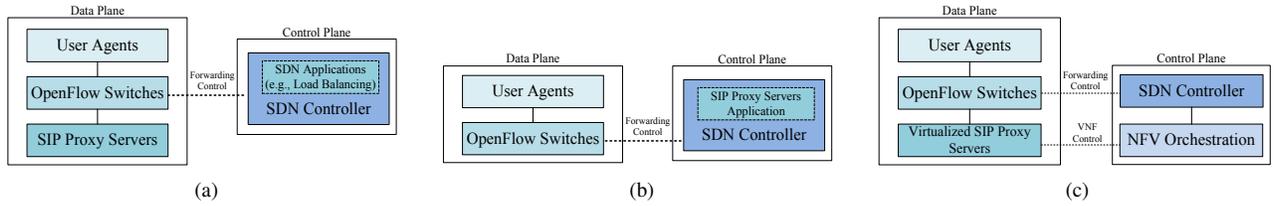

Fig. 6. The architecture of the triple approaches toward softwarization of SIP networks (a) OpenSIP$^{\text{Partial}}$ (b) OpenSIP$^{\text{Full}}$ (c) OpenSIP$^{\text{NFV+}}$

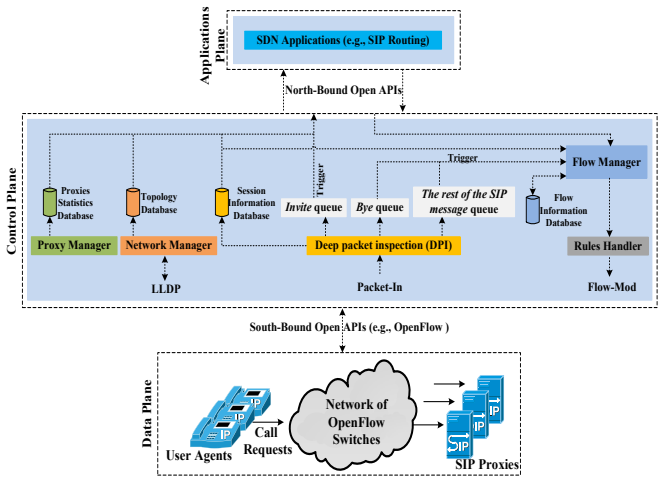

Fig. 7. The architecture of the OpenSIP$^{\text{Partial}}$ controller

### A. Partial SDN-based Architecture (OpenSIP$^{\text{Partial}}$)

A typical method for distributing the load among SIP proxies is employing load balancers. Load balancers may limit network performance and impact the complexity of SIP networks. In contrast to this method, the present paper proposes a switch and OpenFlow controller framework as a first approach. Here, it is assumed that there is an SDN controller in each SIP domain whose responsibility is to manage the OpenFlow switches of that domain. The management operates in such a way that requests are effectively distributed among the proxies of that domain. Load distribution is performed according to the remaining capacity of the proxies and is, in fact, a smart routing process. As seen in Fig. 6(a), the current study attempts to achieve the necessary skill for the effective routing of call requests between UAs and proxies in each domain. This is done by either 1) changing traditional switches with those of the OpenFlow in the data plane or 2) inserting a control plane with centralized logic for the switches.

Fig. 7 shows the controller architecture used in this framework. Given that SDN only concentrates on layer 2-4 information, the DPI module is employed for detecting SIP messages and extracting information. When receiving SIP messages, the switches of this flow encapsulate the message in a `Packet-In` message and send it to the controller for determining their destination. By inspecting the messages, the DPI module classifies and divides them into three queues, namely: *Invite*, *Bye*, and *the Rest of SIP Messages*. Moreover, the session information, including *Call-ID*, *To*, *From*, *Via*, etc., are extracted from the messages and stored in the *Session Information Database*. As mentioned in Section II-A, routing is performed for the `Invite` messages. The other messages and corresponding responses follow the same route (or not, depending on the configuration of the involved proxies). The *Routing Application* is in charge of finding the proxy with the minimum load and selecting the optimal route to reach the destination proxy. In two databases called the *Topology Database* and *Proxies Statistics Database*, *Network Manager* modules, together with the *Proxy Manager*, preserve network topology and proxy information in their respective order. With the Link Layer Discovery Protocol (LLDP)[2] [16] and the *counters* field, the network topology and proxy load are derived, respectively. The *Routing Application* utilizes all three databases to route `Invite` messages. To this end and for the first `Invite` message, the SIP proxy with the minimum load is first chosen. Next, the path between the switches for this message is determined (the UA source and the previously chosen destination proxy) by the powerful *Dijkstra's* algorithm (the shortest path). The task of the *Flow Manager* is to manage sessions according to the *Session Information Database* and the outcome of the *Routing Application*. The information relating to each path and each session's server is stored in the *Flow Information Database* and sent by the *Rules Handler* module to the switches in the `Flow-Mod` messages. This allows the OpenFlow rules to be installed on the switches. The `non-Invite` messages in a specific session follow the route and the chosen server for the `Invite` messages of that session. If the message is `Bye`, the session has to be terminated and the information for that session should be eliminated by the *Flow Manager*. SIP messages reach the destination proxy in five steps (Fig. 8): (1) They may arrive at a switch where no corresponding rule has been installed in the table. Therefore, the switch cannot forward the packet on its own; (2) The switch notifies the controller about the message; (3) The controller identifies the proper path and server for the message; (4) The controller installs the appropriate rules in all switches along the path; (5) SIP message of that session can be forwarded to their destination.

---

[2]The *Network Manager* Module is in charge of sending LLDP packets to all the connected switches through packet out messages. These messages instruct the switches to send LLDP packets to all ports. Once a switch receives the `Packet-Out` message, the LLDP packets are sent out among all the ports. If the neighbor device is an OpenFlow switch, it will perform a flow lookup. Since the switch does not have a flow entry for this LLDP message, it will send this packet to the controller by means of a `Packet-In` message. When the controller receives the `Packet-In`, it analyses the packet and creates a connection in its discovery table for the two switches. All remaining switches in the network will similarly send a packet into the controller, which would create a complete network topology. LLDP messages are periodically exchanged and packets are brought to the controller when links go up/down, or new links are added/removed [16]. Information on switches and links are maintained in the *Topology Database*.

Since the switches are still not able to detect SIP messages, all these messages must be sent to the controller. This causes switch-to-controller round-trip delays. To remedy this problem, the present work effectively extends the structure of OpenFlow switches (Fig. 9). For packet processing, this switch runs software DPI and identifies the attributes of the flows, which are:

- Application ID: SIP, SMTP, Skype, etc.
- Extracted metadata: SIP caller/callee, Call-ID, From, To, Via, etc.
- Computed metadata: Delay, jitter, response time, etc.

After entering one of the switch ports, the attributes of a SIP message are identified by packet processing and are matched with the *match* fields of the flow table. Given that the *match* field in the standard OpenFlow features header information up to layer 4, the OpenFlow rule structure in the flow table are modified according to Fig. 10. In this case, the *match* field contains the header information up to the application layer. The *action* field keeps the operations that can be applied on SIP messages, while the *stats* field holds the computed SIP metadata for smart awareness of the proxy status and network. Finally, if no rule is detected for a SIP message, it is handed to the API part for transmission to the controller. If a rule is derived, the corresponding action is performed (e.g. forward SIP messages to ports). With this mechanism, there is no need to transmit all SIP messages to the controller. This mechanism gradually reduces the number of messages exchanged between the switches and controller.

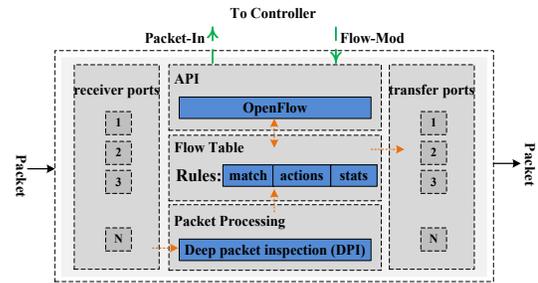

Fig. 9. The structure of an extended OpenFlow switch aware of layer 7

### B. Full SDN-based Architecture (OpenSIP$^{Full}$)

In this approach, the SIP proxy functionalities are decoupled from the infrastructure layer and move to a programmable software controller (Fig. 6(b)). The proposed controller architecture for this approach is shown in Fig. 11. The most vital function of SIP proxies is routing, which forms the application plane. The SIP proxies and SIP registrar structure have been effectively added to the control plane. The data plane merely contains user equipment and extended OpenFlow switches that reduce the management complexities and expenses in this plane. The *DPI* module in this controller classifies the received SIP messages into request and response categories. The *Registrar Manager* module is responsible for registering user information. The *Network Manager* is aware of the switch and link statuses. Routing `Invite` requests is performed with the *Routing Application*. Again, the path of the received `Invite` message between the switches (source and destination are UAC and UAS, respectively) is determined by using *Dijkstra's* algorithm. The *Flow Manager* module is responsible for the management of the sessions. By receiving the `Invite` message path information from the *Routing Application* and accessing the *Session Information Database*, the *Flow Manager* module is capable of routing the remainder of the session (there is mapping between `Invite`(s) and flow(s)). In the SIP network, UAS produces some of the responses (e.g. `180 Ringing`) and proxy generates the other messages (e.g. `100 Trying`) with respect to the request message. The *Make an Appropriate Response Manager* module is responsible for generating the response messages of the second type (e.g. `100 Trying`) in the `Packet-Out` message.

Fig. 12 presents messages related to user information registration, initiation and the tearing down of a session by this approach. Contrary to Fig. 2, there is no SIP proxy equipment in this figure. In the registration phase, the `Register` message is sent to the controller via a switch and the `Packet-In` message for user information registration. For the acknowledgment, the controller sends a `Packet-Out` message to the switch that holds a `200 Ok` message. In the initiation phase, via the `Packet-In` message, the first switch sends a copy of the received `Invite` message to the controller.

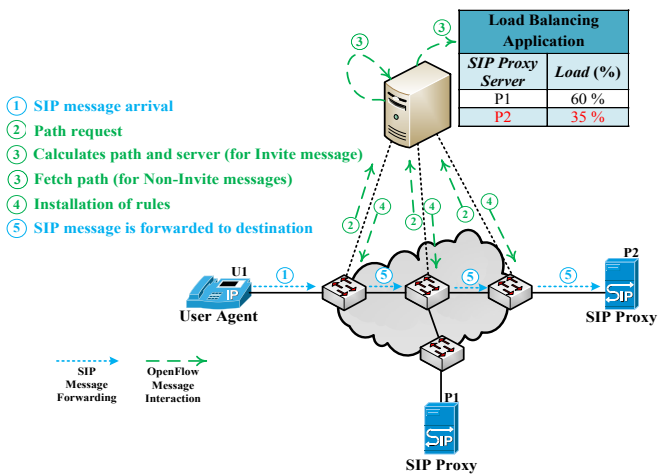

① SIP message arrival
② Path request
③ Calculates path and server (for Invite message)
④ Fetch path (for Non-Invite messages)
④ Installation of rules
⑤ SIP message is forwarded to destination

Fig. 8. Flow management

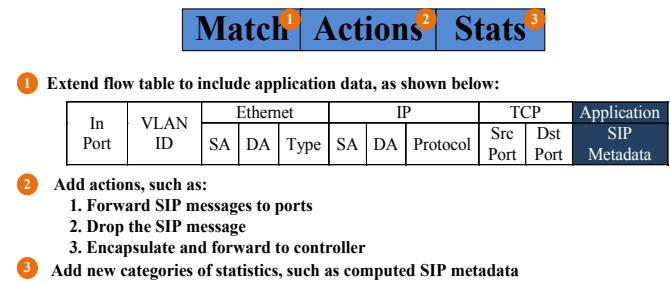

❶ Extend flow table to include application data, as shown below:

| In Port | VLAN ID | Ethernet | | | IP | | | TCP | | Application |
|---------|---------|----|----|------|----|----|----------|-----|-----|-------------|
| | | SA | DA | Type | SA | DA | Protocol | Src Port | Dst Port | SIP Metadata |

❷ Add actions, such as:
   1. Forward SIP messages to ports
   2. Drop the SIP message
   3. Encapsulate and forward to controller

❸ Add new categories of statistics, such as computed SIP metadata

Fig. 10. Extension of the OpenFlow protocol to support SIP

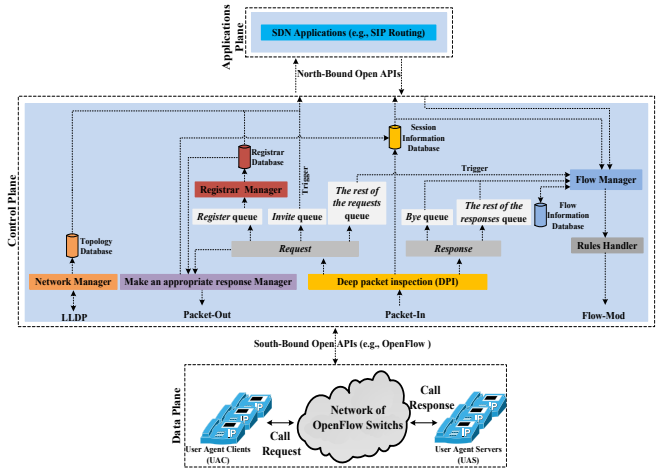

Fig. 11. The architecture of the OpenSIP^Full controller

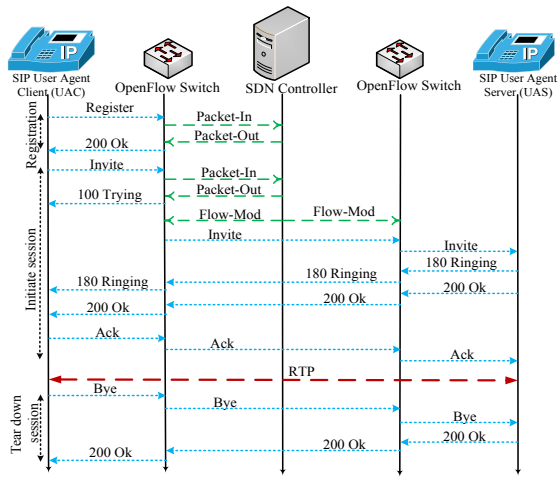

Fig. 12. Advanced SIP call flow

The path of the `Invite` message in the network switches to its destination (UAS) is also determined by the controller and is given to corresponding switches using the `Flow-Mod` message. At this point, the `Invite` message is sent to the destination via the switches. As the rules for this session are already installed on the switches, the remainder of the request and response messages are exchanged with no controller supervision.

### C. Software-defined NFV-based Architecture (OpenSIP^NFV+)

Leveraging cloud technologies (e.g. hardware virtualization) as key enabler is an emerging concept (called NFV), and is currently receiving great attention. It is worth mentioning that both Cloud computing and NFV are more than simple concepts, that is they are being deployed on a commercial-scale today. The main goal behind the NFV is to enable the consolidation and sharing of various software-based, virtualized, networking resources, running on commodity hardware infrastructures. NFV could also be used to improve the performance of SIP networks. As mentioned earlier, the efficiency of SIP proxies is highly affected by their processing resources. This is where NFV comes to the rescue as it helps to eradicate the limitations of hardware-centric SIP proxy and reduces the possible occurrences of overload. This is the reason why *Software-Defined Network Function Virtualization* OpenSIP^NFV+ is proposed.

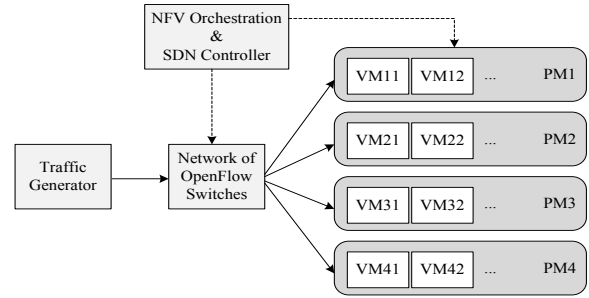

Fig. 13. Load distribution between virtual SIP proxies

Software-defined NFV leverages network virtualization and logically centralized intelligence so as to minimize costs, and to maximize the utilization of network resources. In OpenSIP^NFV+, VNFs in the data plane are virtual SIP proxies. Also, all networking resources (such as switches) are under the control of the SDN controller, while all computing resources (such as virtual SIP proxies) are under the control of NFVO (Fig. 13). As seen in Fig. 6(c), hypervisors run on servers to support the Virtual Machines (VMs) that implement SIP Proxies. This platform allows customizable and programmable virtualized SIP proxies that run as software within virtual machines. Therefore, SIP proxies are delivered to the network operator as pieces of pure software. Note that *on demand virtual resources* from Cloud computing can be leased and released, which is a promising paradigm that promotes a computing-as-a-service model. The SDN controller and the NFV orchestration system make up the logical control plane. NFV orchestration manages virtual SIP proxies. By using standard interfaces, the SDN controller controls NFV orchestration. After the allocation of suitable SIP Proxies to virtual machines via NFV orchestration, the SDN controller initiates efficient call request steering through OpenFlow switches and with the *Dijkstra* algorithm. To this end, the controller and switch architecture are the same as those of the first approach, with the exception that the application protocol is responsible for autoscaling VMs based on the input load. Later the proper path for delivering `Invite` messages is determined. Autoscaling is a technique for the dynamic adjustment of resources with respect to demand [14]. For this approach, we utilize horizontal scaling, which increases the scalability of the network and overall reduces the possibility of overload.

According to Fig. 13, based on the demand, some VMs are hosted by Physical Machines (PMs). Moreover, a SIP proxy is implemented on each VM. The load is first sent to VM11. The scale out/in criteria is that the consumed resources of all VMs exceed 90% or fall below 10% of the resources, respectively. For scale out, NFV orchestration launches VM21 on PM2 and, for scale in, the load is allocated to previous VMs with sufficient resources.

As a conclusion, Table I represents a comparison between the various version of OpenSIP.

### D. OpenSIP in IMS and Multi-domain Networks

Integrating the OpenSIP to the IP Multimedia Subsystem (IMS) network can provide better resource control and advanced QoS. Traditionally, service subscription or modification on the network bandwidth needs certain time of handling to take effect. With the incorporation of OpenSIP, it becomes real-time. In this regard, the call is routed to the IMS core and triggered to the SIP application servers (AS) responsible for the service of network resource control. After proper authorization and authentication, the SIP AS can use the OpenSIP controller's north-bound API to modify the subscribed network resources according to the calling party's requirement. In a multi-domain SIP network, the logically centralized OpenSIP controller consists of multiple distributed controllers working together, which can benefit from the scalability of the distributed architecture. Each controller only handles the local area switches in its domain. OpenSIP controllers can be horizontal or vertical (Fig. 14(a) and (b)). In the vertical approach, a master OpenSIP controller is on top of the individual network controllers. Master OpenSIP controller has a global view of the network across all connected domains. It can orchestrate the configuration in each connected domain. In the horizontal approach, the OpenSIP controllers establish peer-to-peer communication. Each controller can request information or connections from controllers of other domains in the network.

TABLE I
THE LOCATION OF ARCHITECTURAL MODULES IN DIFFERENT APPROACHES

| Approaches | Conventional | OpenSIP[Partial] | OpenSIP[Full] | OpenSIP[NFV+] |
|---|---|---|---|---|
| routing | data plane | application plane | application plane | application plane |
| DPI | - | control plane | control plane | control plane |
| proxy | data plane | data plane | control plane | data plane |
| registrar | data plane | data plane | control plane | data plane |

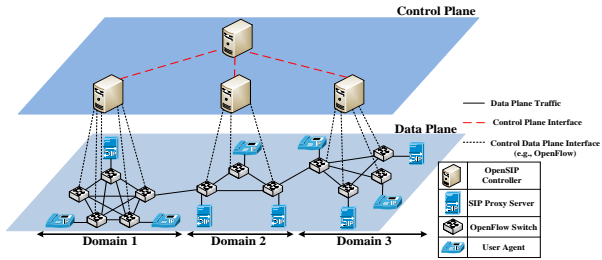

(a) Vertical approach

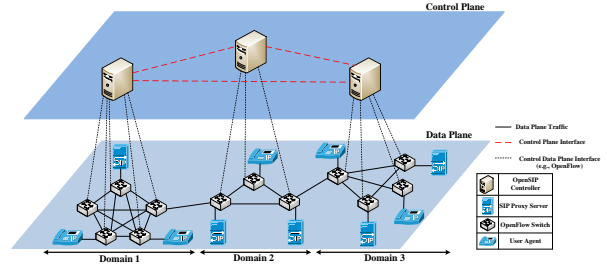

(b) Horizontal approach

Fig. 14. OpenSIP support of multi-domain networks

## IV. IMPLEMENTATION AND PERFORMANCE EVALUATION

This section first provides detail on the implementation of the system. Then, the performance and the results of the three proposed approaches are evaluated and presented in consecutive sections. *Open vSwitch v2.4.1* and *Floodlight v1.2* are employed to implement the OpenFlow switch and controller and to extend them in accordance with what was described in the previous section ($\sim$ 2400 lines of Java code). Floodlight is a multi-threaded Java-based controller that utilizes the Netty framework. Open vSwitch is a software implementation of a virtual multilayer network switch, which is designed to enable effective network automation through programmable extensions while supporting the OpenFlow protocol. In addition, the application plane is implemented as a module running on the controller. *nDPI engine* is used to implement the DPI module. nDPI is an open source deep packet inspection software toolkit. The open source *Kamailio v4.3.6* software implements the SIP proxies and the open source *SIPp* software implements the UAs and injects SIP traffic. *OProfile* software is utilized to measure CPU and memory usages. When necessary, to send packets at a fixed rate, the background traffic between proxies is generated by *iperf*. To sniff and capture the network packets, *Wireshark* is used. Each experiment is run three times and the average is taken as the result.

### A. Experimental Results for OpenSIP[Partial]

*1) Constant Load:* To evaluate the performance of the OpenSIP[Partial] approach, the topology shown in Fig. 15 is employed.

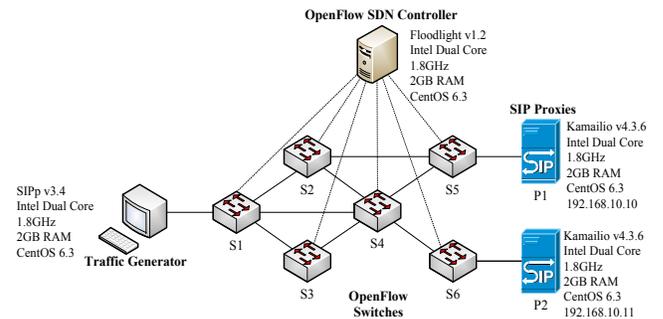

Fig. 15. The testbed for implementation of OpenSIP[Partial]

This topology includes two SIP proxies, six OpenFlow switches, and one controller. For implementation of the switches, Open vSwitch is installed on six virtual machines. Each network link has a bandwidth of 10 Mbps. The first experiment entails two scenarios with different background traffic. In Scenario 1, the background traffic of each proxy is equally 500 packets. On the other hand, in Scenario 2 the background traffic of proxy 1 (P1) is 1000 and that of Proxy 2 (P2) is 500 packets. Then, the constant offered load is injected through the traffic generator to the system for 100 seconds at the rate of 1500 cps (call per second).

The performance of OpenSIP[Partial] with that of the Round-robin and Random proxy selection strategies is compared. To be more precise, *Routing Application* is composed of two parts: *SIP proxy selection* and *Path selection*. In this experiment, three algorithms are used in order to implement the *SIP proxy selection*: 1. The proposed method of the present paper based on the *counters* field and designed modules such as *Proxy Manager* and *Network Manager*, 2. Round-robin, and 3. Random. So according to these algorithms, the SIP proxy (with the minimum load in proposed method case) is first chosen. Then, all three algorithms exploit the *Dijkstra's* algorithm as *Path selection* algorithm (the shortest path). Fig. 16 presents the performance of the SIP proxies. The evaluation criteria include proxy throughput (the number of serviced calls by proxies per unit of time), the proxies' average response time (the time between sending the `Invite` from the UA and receiving the `200 Ok` from the SIP proxy), and the consumption of resources by the proxies. The objective is to achieve the maximum throughput and the minimum latency with respect to resources. Compared to the Round-robin and Random approaches, the OpenSIP$^{Partial}$ approach reaches better results in both scenarios. In addition, the OpenSIP results are similar in both scenarios, yet the results of the Round-robin and Random approaches decline in Scenario 2. The different background traffic of the proxies in Scenario 2 aggravate the careless load distribution of these two methods over time. As shown in Fig. 16.(a) and (b), it is clear that the OpenSIP$^{Partial}$ approach obtains a throughput closest to the offered load in both scenarios. The OpenSIP$^{Partial}$ approach is able to sufficiently approximate the proxy loads using the stats field. Despite the increased resource consumption of the Round-robin and Random approaches compared with the OpenSIP$^{Partial}$ approach (Fig. 16(e) to (h)), their average response time is higher (Fig. 16(c) and (d)). Note that resource consumption by both proxies in the OpenSIP$^{Partial}$ approach is nearly equal, indicating that OpenSIP has a careful and fair load distribution.

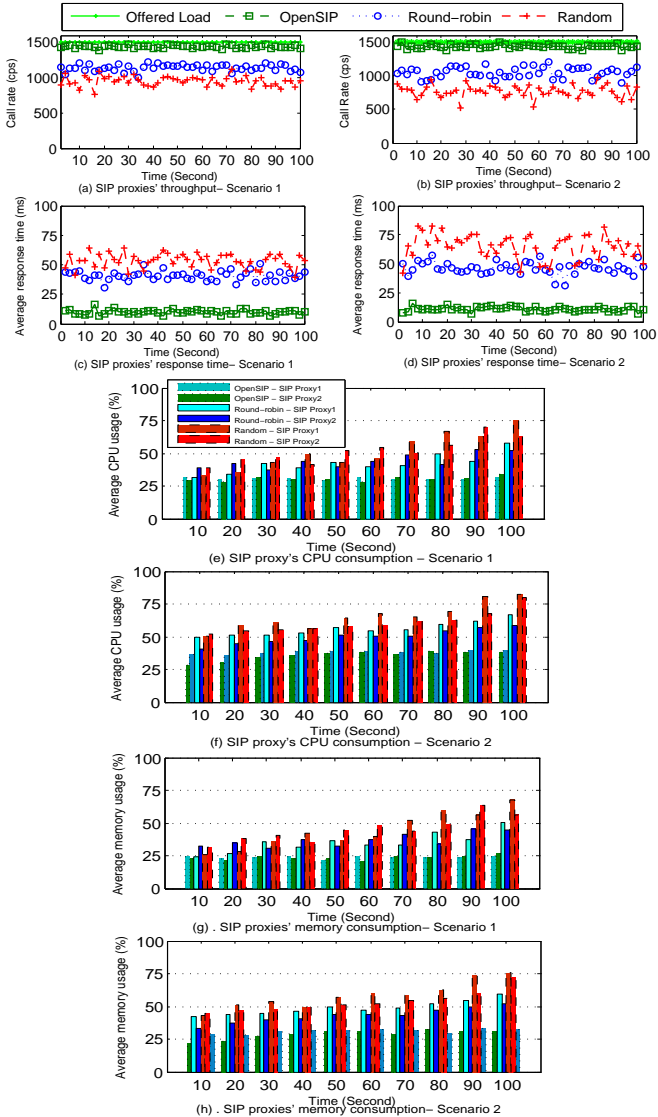

Fig. 16. Comparison of SIP proxies' performance in the two scenarios

The following focuses on OpenSIP$^{Partial}$ results. Each proxy's throughput in the OpenSIP$^{Partial}$ approach is shown in Fig. 17. In

Scenario 1, both proxies have an approximately equal throughput ($\sim$ 748 cps). In Scenario 2, the throughput of P2 is ($\sim$ 997 cps), almost double that of P1 ($\sim$ 501 cps). The reason for this difference is that the background traffic of P1 is half of P2's in this scenario. Therefore, more load should be sent to P2. As shown in Fig. 16(a) and (b), the total throughput of the two proxies is very close to the offered load. Note that the offered load is distributed fairly between the two proxies and in accordance with their available capacity in the OpenSIP$^{Partial}$ approach.

Fig. 18 provides the link utilization at the 50$^{th}$ second in the OpenSIP$^{Partial}$ approach. This is indicative of the routing of requests to P1 and P2. For this purpose, the OpenSIP uses the shortest possible routes of <S1, S2, S5> and <S1, S4, S6>, to send requests to P1 and P2, respectively. In Scenario 2, the links along the routes leading to P2 are double used the links along the routes leading to P1. As mentioned before, in Scenario 2, the background traffic of P2 is half of that of P1. This causes equal resource consumption in both the P1 and P2 proxies (Fig. 16(e) to (h)).

Fig. 19 shows the performance of the OpenSIP$^{Partial}$ controller. The controller throughput indicates the amount of serviced flow per unit of time. The average response time represents the time between the `Packet-In` is sent from the switch and the `Flow-Mod` is received from the controller. As shown in Fig. 19(a) and (b), the performance of the OpenSIP$^{Partial}$ controller is independent of the scenario; that is, its average throughput and average response time are approximately 1450 fps and 7 ms, respectively . This indicates that the OpenSIP$^{Partial}$ controller can achieve a high throughput with a very low latency and without its modules overloading the resources (Fig. 19(c) and (d)). The controllers' resource consumption is lower than the proxies' (compare Fig. 16 and Fig. 19), because the latter are also responsible for initiating and terminating all calls while the controllers are only responsible for managing a limited number of switches. This can also be inferred from Fig. 20. This figure confirms that the proxies' number of processed packets per unit of time is almost seven times that of the controller's. As shown in Fig. 2, seven messages are involved during the SIP call flow (and in the worst case traverse all SIP proxies), while switch installation rules by the controller begin with the `Packet-In` message and end with the `Flow-Mode` message. Therefore, the possibility of overloading the controller is much lower than the proxies' chances, and so a controller will never become a bottleneck. Moreover, with the equitable and fair distribution of loads by OpenSIP, proxies will also never have to deal with an overload.

Another situation which may cause a bottleneck for the OpenSIP$^{Partial}$ controller is the sudden failure of network components and a sudden reduction of its capacity. This failure may impose the failed proxy's load on the other proxies. To test such a situation, in Scenario 1, P1 fails at the 20$^{th}$ second and resets up at the 40$^{th}$ second.

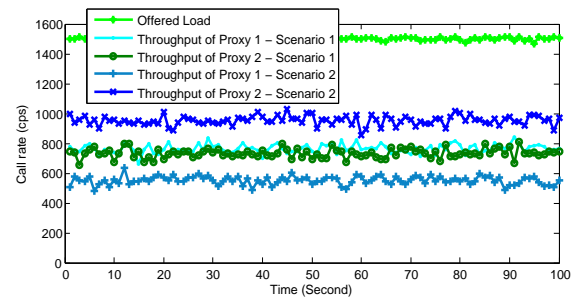

Fig. 17. Fairness analysis in OpenSIP$^{Partial}$

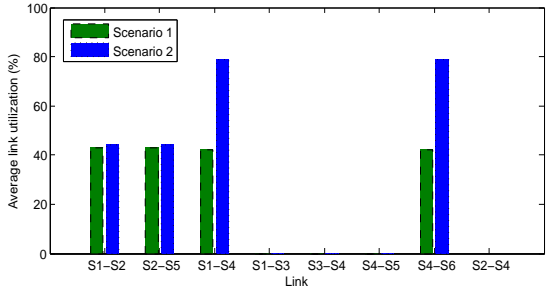

Fig. 18. Link utilization observed at t = 50s in OpenSIP$^{Partial}$

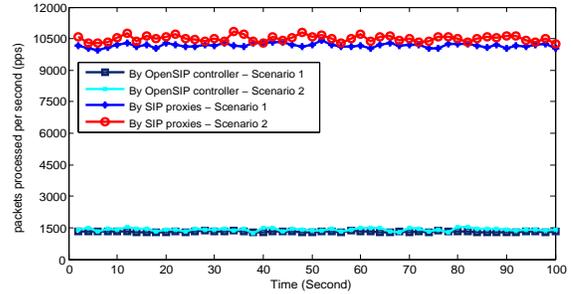

Fig. 20. Packets processed per second in OpenSIP$^{Partial}$

Likewise, Scenario 2, P1 fails at the 60$^{th}$ second and resets up at the 80$^{th}$ second. Fig. 21 shows the OpenSIP$^{Partial}$ performance in these conditions. In Scenario 1, from the 20$^{th}$ second to the 40$^{th}$ second, the OpenSIP$^{Partial}$ controller can send the whole load to P2 (Fig. 21(a)). This indicates that, despite the sudden failure of P1, the OpenSIP speed can still maintain the throughput of the entire system close to the offered load. P2's resource consumption also increases during this period (Fig. 21(b) and (c)). In addition, the resource consumption of the OpenSIP$^{Partial}$ controller grows during this period. Under normal circumstances, P1 and P2's resource consumptions are almost equal, while that of the OpenSIP$^{Partial}$ controller is lower (Fig. 21, (b) and (c)). In Scenario 2, the same process is repeated from the 60$^{th}$ second to the 80$^{th}$ second, but the difference is that the P1 and P2 loads are unequal. However, the OpenSIP$^{Partial}$ can send the whole load to P2 at the 60$^{th}$ second and then distribute it again between P1 and P2 at the 80$^{th}$ second.

*2) Variable load:* In the previous subsection, the constant load of 1500 cps is injected into the system. However, in this subsection, the performance of the OpenSIP$^{partial}$ is evaluated under a variable load. In the previous subsection, the P1 and P2 proxies did not deal with an overload since their capacity was more than 1500 cps and so a lack of resources was not encountered. On the contrary, in this test we seek to evaluate the performance of OpenSIP$^{Partial}$ under overload condition. Fig. 22 presents the performance evaluation results. The offered load is initiated from 1500 cps and increases to 6000 cps in four stages (until the 400$^{th}$ second). Then with a sudden drop at the 400$^{th}$ second, its offered load returns to 1500 cps. Next with a sudden spurt at the 500$^{th}$ second, the load reaches 6000 cps and finally drops back to 3000 cps during the last 100 seconds. Up to the 200$^{th}$ second, the throughput of the proxies and controller are very close to the offered load. At the 200$^{th}$ second, the proxies are overloaded and this will continue up to the 400$^{th}$ second. During these 200 seconds, the average throughput of the proxies is approximately 3000 cps and the remainder of the offered load is rejected by the proxies (rejection rate). The proxies' overload occurs due to a lack of resources, especially CPU (Fig. 22(c) and (d)). For example, Fig. 22(c) shows the proxies' CPU saturation at the 250$^{th}$, 350$^{th}$ and 550$^{th}$ second, respectively. Usually, with the saturation of CPU, throughput approaches zero. However the OpenSIP$^{Partial}$ overcomes the sharp drop in throughput during the overload status, which allows it to use the proxies maximum capacity ($\sim$ 3000 cps). Unlike the proxies, the OpenSIP$^{Partial}$ controller is not overloaded even at the peak offered load and its throughput is always synchronized with the offered load. For example, the average throughput of OpenSIP$^{Partial}$ between 300 to 400 seconds is approximately 5978 fps. Proportional to the throughput, proxy response time also varies (Fig. 22(b)). Upon overload, response time increases several times as well. After the offered load returns to an amount lower than the proxies' capacity (that is, back to a situation without the overload), an overload control algorithm may not be able to return the throughput to a normal

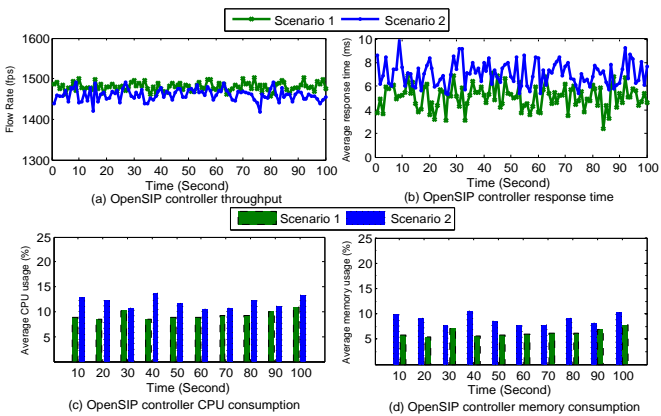

Fig. 19. Comparison of the OpenSIP$^{Partial}$ controller's performance in the two scenarios

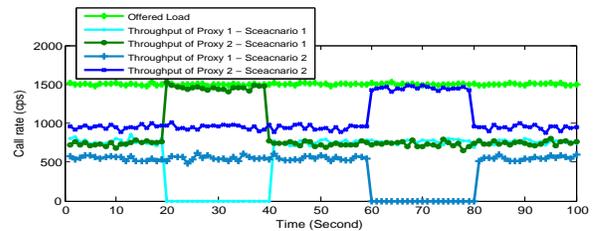

(a) Throughput

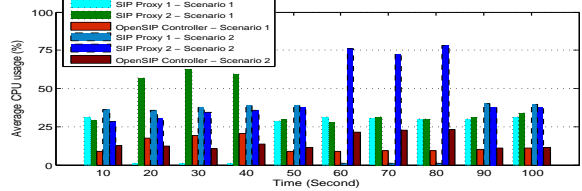

(b) CPU Consumption

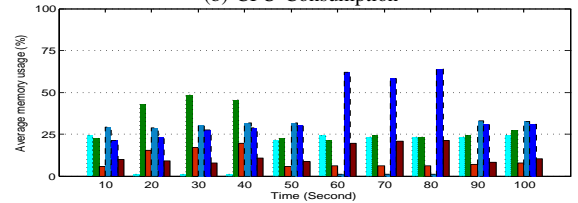

(c) Memory Consumption

Fig. 21. OpenSIP$^{Partial}$ performance at the time of P1 failure

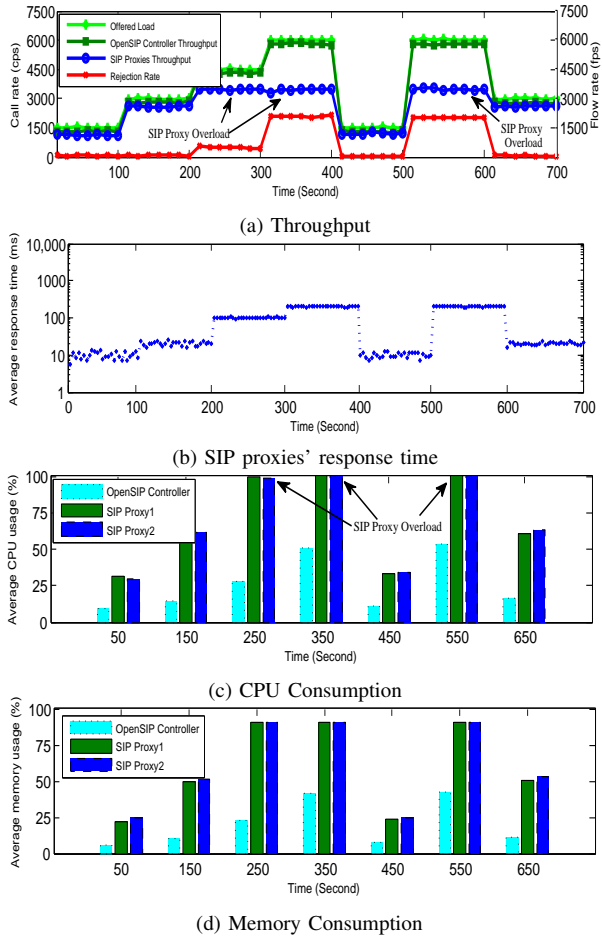

(a) Throughput

(b) SIP proxies' response time

(c) CPU Consumption

(d) Memory Consumption

Fig. 22. OpenSIP^Partial performance over time in various offered loads

status. However, at the end of the overload (at the 400th second), the OpenSIP^Partial can bring the system throughput close to the offered load by a proper distribution of the load. Unlike the gradual increase of the offered load in the first 400 seconds, a flash crowd occurs at the 500th second. At the same time, the OpenSIP^Partial can again manage to maximize the proxies' throughput despite the total occupation of their CPU (Fig. 22(c)). A flash crowd occurs when a large number of user agents simultaneously make call requests. For example, on holidays, numerous calls are made within short intervals, which imposes a heavy load on the network. The proxies' high throughput during sudden fluctuations of the offered load points to the stability of the system. Finally, in the last 100 seconds, the rejection rate reaches the minimum while the throughput attains the maximum, and the response time reaches approximately 10 ms. The point to note here is that, during the period between 200 to 400 or 500 to 600 seconds when the input load exceeds the network resources, a throughput close to the offered load can be achieved by increasing proxy resources and overcoming their hardware limitations. This condition is the motivation behind the third proposed approach, which is entirely based on NFV, and whose performance results are presented in the following subsection.

*3) Comparison with Other Approaches:* In this subsection, the performance of the OpenSIP^Partial is compared with two other well-known approaches given in [17] and [18]. For this purpose, as shown in Fig. 23, two testbeds are prepared, whose details are provided in Table II.

In [17], using a load balancer, a TLWL algorithm routes a new call request to the SIP proxy with the least load. The counters in the TLWL algorithm specify the weighted total of the transactions assigned to each proxy. A new call is assigned to the SIP proxy with the lowest counter. In [18], to distribute the load with a load balancer, the HWAR and FWAR algorithms are proposed, which use the window and proxy response time to estimate the proxies' load. In the FWAR algorithm, windows sizes are fixed, whereas in HWAR a history of the response times is kept in the windows with the help of a mathematical model. Fig. 24 shows achievable peak throughput of the proxies using each method. As is evident, OpenSIP^Partial can achieve the highest throughput (5856 cps), while the HWAR is only able to handle up to 4940 cps. Fig. 25 also lists the average response time for proxies in the four methods. It is clear that the average response time of OpenSIP^Partial has a linear growth, while it is exponential for the other methods.

Table III presents the proxies' CPU consumption along with their means ($\mu$) and standard deviations ($\sigma$). It is evident that by

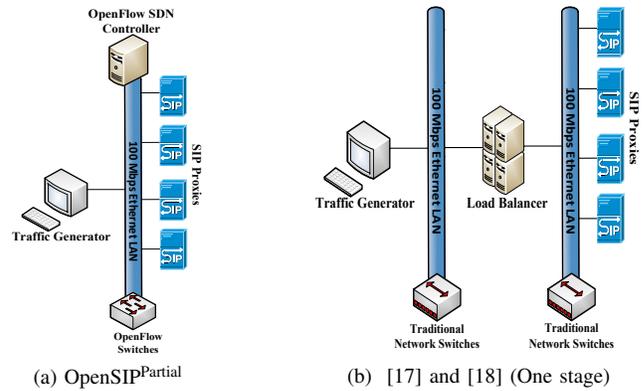

(a) OpenSIP^Partial

(b) [17] and [18] (One stage)

Fig. 23. Implementation testbeds prepared to compare OpenSIP^Partial with [17] and [18]

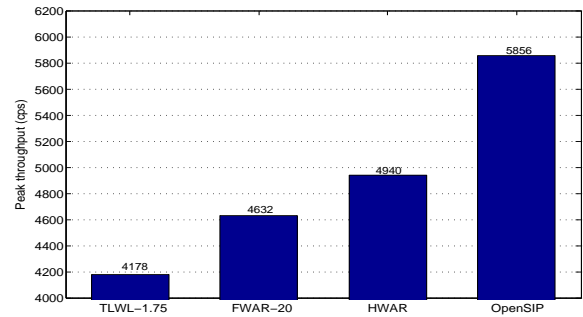

Fig. 24. SIP proxies' peak throughput in various methods with eight SIP proxies

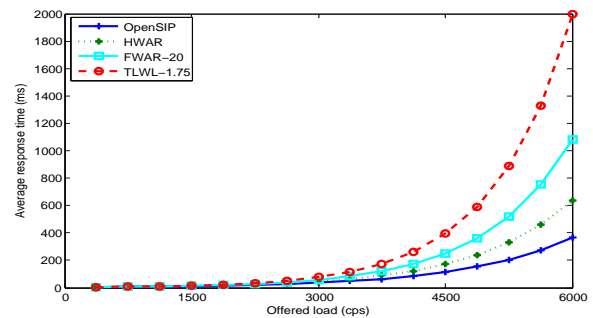

Fig. 25. SIP proxies' response time in various methods with eight SIP proxies

TABLE II
Testbed characteristics for the third experiment

|  | SIP Proxies | SDN Controller | OpenFlow Switches | Load Balancer | Traffic Generator (UAs) |
|---|---|---|---|---|---|
| Software | Kamailio v4.3.6 | Floodlight v1.2 | Open vSwitch v2.4.1 | Asterisk v13.10.0 | SIPp v3.4 |
| Quantity | 8 | 1 | 16 | 1 | 1 |
| CPU | Intel Dual Core 1.8GHz | Intel Xeon E5645 2.4GHz* | Intel Dual Core 1.8GHz | Intel Xeon E5645 2.4GHz | Intel Dual Core 1.8GHz |
| RAM | 2 GB | 4 GB | 1 GB | 4 GB | 2 GB |
| Operating System | CentOS v7.2 | Linux kernel v3.10 | Linux kernel v3.10 | CentOS v7.2 | Red Hat v6 |

* 6 cores - 12 threads

increasing the offered load, the proxies' CPU usage also grows. This table offers an insightful view of the manner of load distribution among the eight proxies using the four methods. Compared with the other methods, the standard deviation of CPU consumption in the OpenSIP$^{Partial}$ method is lower. This indicates a uniform load distribution among proxies. Moreover, average CPU consumption is lower for OpenSIP$^{Partial}$ except for offered load of 1500 cps, which represents an efficient use of resources. When the offered load is equal to 1500 cps, both the FWAR and TLWL methods have a lower $\mu$, but a higher $\sigma$. For example, CPU consumption of proxy 3 in TLWL is approximately 38%, while it is 4.4% for proxy 8 (a significant difference!). The memory results are similar though they, they will not be presented here.

It is also worth mentioning that OpenSIP$^{Partial}$ scales well (at least up to 12 proxies) as presented by Fig. 26.

Fig. 27 shows the proxies' peak throughput versus an increase in the number of the OpenFlow switches. The base of the Open-SIP is Floodlight, which is a multi-threaded controller. For multi-threaded controllers, adding more switches leads to better utilization of available CPU cores. Thus, throughput increases until the number of connected switches is larger than the number of threads. Since the OpenSIP processor has 6 cores and 12 threads (Table II), increasing the number of switches to 12 boosts the proxy throughput.

In Fig. 28, the performance of the load balancer and controller entities are compared with each other. To do so, the offered load of 1000 cps to 10000 cps is injected into the entities in both testbeds (Fig. 23). The OpenSIP$^{Partial}$ controller achieves a throughput of 6650 fps, while the best load balancer belonging to the HWAR algorithm can not achieve a better throughput than 5040 cps. This is because to estimate the proxies' loads using the HWAR, FWAR, and TLWL, information must be stored and processed in the load balancer. It is also necessary to note that, according to Table II, the hardware specifications of both entities are similar.

In addition, we have extended the comparison testbed to a two stage with ECMP testbed, and repeated the experiment. Table IV illustrates the results. Despite performance improvements of TLWL in the new testbed, the results still show still shows low yields compared to the

OpenSIP$^{Partial}$, which is the result of different architectures. The high performance of OpenSIP compared to TLWL comes from the efficient and integrated framework based on SDN concept, which decouples the network control from data forwarding by direct programming. With its inherent decoupling of control from data plane, SDN presents a greater control of network through programming.

### B. Experimental Results for OpenSIP$^{Full}$

To evaluate the performance of the OpenSIP$^{Full}$ approach, the topology shown in Fig. 12 is used. This includes the SDN controller (namely, the OpenSIP$^{Full}$ controller), OpenFlow switches, and user agents according to the specifications given in Table II. Fig. 29 shows a wireshark capture of messages exchanged for user registration, session initiation, media exchange and session termination. As is clear, a VoIP session can be established with no physical SIP proxy and with a minimum number of messages exchanged. The point to note is that mapping occurs between the SIP and the OpenFlow messages. As seen in this figure, after receiving the `Register` message from the user agent (192.168.1.100), the OpenFlow switch (192.168.1.101) sends this to the OpenSIP$^{Full}$ controller (192.168.1.102) in the form of a `Packet-In` message. The controller registers the user information and sends its acknowledgment as a `Packet-Out` message to the switch. This message contains `200 Ok` message for the user agent.

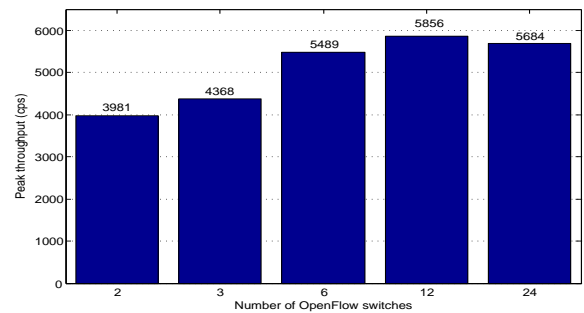

Fig. 27. SIP proxy peak throughput vs. the number of OpenFlow Switches (with eight SIP proxies in OpenSIP$^{Partial}$)

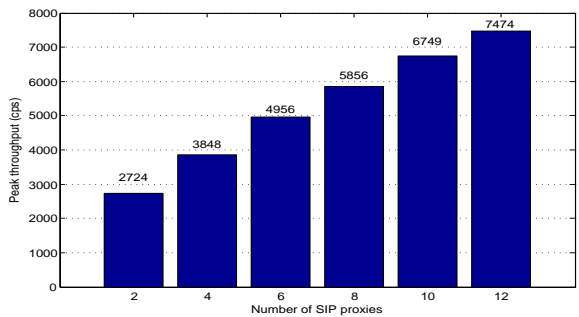

Fig. 26. SIP proxy peak throughput vs. the number of SIP proxies (with sixteen OpenFlow Switches in OpenSIP$^{Partial}$)

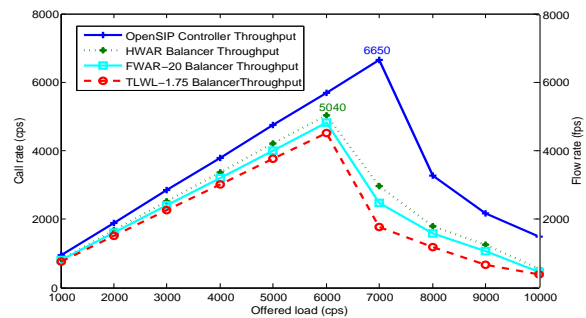

Fig. 28. OpenSIP$^{Partial}$ controller throughput vs. load balancer throughput

TABLE III
Proxies' CPU consumption, Mean ($\mu$) and Standard deviation ($\sigma$)

| Offered Load (calls/sec) ~ | | | | | 1500 cps | | | | 3000 cps | | | | 6000 cps | | | |
|---|---|---|---|---|---|---|---|---|---|---|---|---|---|---|---|
| CPU Usage(%) ~ \ Methods | OpenSIP$^{Partial}$ | HWAR | FWAR-20 | TLWL-1.75 | OpenSIP$^{Partial}$ | HWAR | FWAR-20 | TLWL-1.75 | OpenSIP$^{Partial}$ | HWAR | FWAR-20 | TLWL-1.75 |
| SIP Proxy 1 | 28.28711 | 34.41847 | 11.41962 | 17.6959 | 46.92671 | 50.77466 | 60.41263 | 63.67234 | 87.1522 | 99.4433 | 98.35628 | 100 |
| SIP Proxy 2 | 29.11409 | 26.58059 | 24.59085 | 11.13422 | 53.36199 | 46.76779 | 59.75894 | 62.36888 | 84.70425 | 98.4757 | 100 | 70.52097 |
| SIP Proxy 3 | 26.49163 | 29.97922 | 35.79715 | 37.93151 | 45.36459 | 60.22029 | 47.02127 | 28.92115 | 89.76284 | 91.90538 | 69.40477 | 100 |
| SIP Proxy 4 | 21.10002 | 25.95929 | 24.88893 | 37.12324 | 48.37223 | 55.24889 | 49.48492 | 73.77138 | 86.17214 | 83.34733 | 99.4757 | 98.92849 |
| SIP Proxy 5 | 23.32516 | 26.64898 | 3.990736 | 29.97475 | 45.2882 | 38.25194 | 55.63625 | 56.92173 | 95.58709 | 94.58904 | 100 | 99.1262 |
| SIP Proxy 6 | 24.25989 | 29.18996 | 24.12473 | 15.40299 | 39.17503 | 56.90217 | 53.87371 | 63.66082 | 94.35496 | 95.34923 | 83.92338 | 81.99253 |
| SIP Proxy 7 | 23.84553 | 19.34619 | 33.5041 | 19.89493 | 51.81007 | 51.5295 | 54.52989 | 46.85648 | 87.74447 | 86.6149 | 86.38814 | 99.023 |
| SIP Proxy 8 | 23.51356 | 26.30175 | 32.47848 | 4.4203 | 49.62805 | 37.72943 | 54.14984 | 53.15557 | 90.788 | 90.77296 | 92.1798 | 100 |
| $\mu$ | 24.99212 | 27.30306 | 23.84932 | 21.69723 | 47.49086 | 49.67934 | 54.35843 | 56.16604 | 89.53324 | 97.56223 | 93.80874 | 99.8833 |
| $\sigma$ | 2.550988 | 4.006575 | 10.37422 | 11.39614 | 4.142978 | 7.752312 | 4.262402 | 12.74522 | 3.623025 | 9.082369 | 12.13353 | 14.62698 |

TABLE IV
Comparison between the new results of TLWL in two stage testbed and OpenSIP$^{Partial}$

| Time (sec) | 0 - 100 | 100 - 200 | 200 - 300 | 300 - 400 | 400 - 500 | 500 - 600 |
|---|---|---|---|---|---|---|
| Offered load (calls/sec) ~ | 1000 | 2000 | 3000 | 4000 | 5000 | 6000 |
| SDN-based testbed — OpenSIP$^{Partial}$ throughput (calls/sec) ~ | 988 | 1968 | 2878 | 3879 | 4921 | 5436 |
| SDN-based testbed — OpenSIP$^{Partial}$ average response time (ms) ~ | 9.64 | 16.44 | 27.98 | 51.46 | 81.76 | 99.65 |
| Two stage testbed — TLWL throughput (calls/sec) ~ | 759 | 1108 | 1989 | 2897 | 3789 | 4497 |
| Two stage testbed — TLWL average response time (ms) ~ | 9.98 | 19.64 | 48.55 | 76.55 | 176.64 | 265.28 |

The call request is also sent via an `Invite` message and its route is specified by a `Flow-Mod` message for the OpenFlow switches. Then, the SIP messages are exchanged between the user agents without the need of sending them to the controller.

Fig. 30 reports the OpenSIP$^{Full}$ controller scalability. This controller is able to achieve a maximum throughput of 4897 fps by increasing the offered load to 5000 cps. The comparison of Fig. 28 and Fig. 30 indicates that the OpenSIP$^{Partial}$ achieves a higher throughput than the OpenSIP$^{Full}$ because the OpenSIP$^{Full}$ also has the responsibility of the SIP proxy. Table V shows that up to the offered load of 5000 cps, the OpenSIP$^{Full}$ controller response time is less than 100 ms, but after that it grows quickly.

Fig. 31 provides the CPU profiling results for the OpenSIP$^{Full}$ controller obtained via *Oprfile* for various offered loads. As observed, the *Floodlight Kernel* Module occupies more than half of the CPU of the controller, but the modules designed in this paper use much less CPU. This further proves the scalability of the OpenSIP$^{Full}$. Among these modules, the *DPI* Module has a higher CPU consumption due to the inspection of the received messages up to layer 7. However, the *DPI* engine can improve so as to reduce resource consumption. Fig. 31 also illustrates that, at the offered load of 5000 cps, the CPU of the controller is fully engaged, thus the decreased throughput and increased response time are now justifiable.

### C. Experimental Results for OpenSIP$^{NFV+}$

To evaluate the performance of the OpenSIP$^{NFV+}$ approach, the topology presented in Fig. 13 is employed according to the specifications given in Table VI. Table VII shows that, with the passage of time and the increase of the offered load, the OpenSIP$^{NFV+}$ approach is able to maintain a high throughput by setting the number of VMs by NFV orchestration. The OpenSIP$^{NFV+}$ prevents losses, even in traffics as heavy as 6000 cps from 500 to 600 seconds, and achieves a throughput of 5867 cps with six virtual SIP proxies.

As a conclusion, Table VIII represents a comparison in terms of throughput and delay between the aforementioned methods. As expected, OpenSIPs can achieve a high throughput with a very low delay. OpenSIPs are indebted to the SDN, OpenFlow, and global view of entire SIP network for high efficiency.

### V. Related Work

SIP overload control algorithms are divided into local and distributed. In the local, the overloaded proxy has control over its resource usage [19]. The criteria for identifying overload in these algorithms are queue length and CPU usage. According to these criteria, a set of thresholds is defined that when exceeded, makes the proxy enter overload condition. The proxy then starts to reject call requests. For example in [20], occupancy-based algorithms (OCC) are proposed which employ CPU utilization as a trigger for rejecting calls. Another well-known example of the local overload control

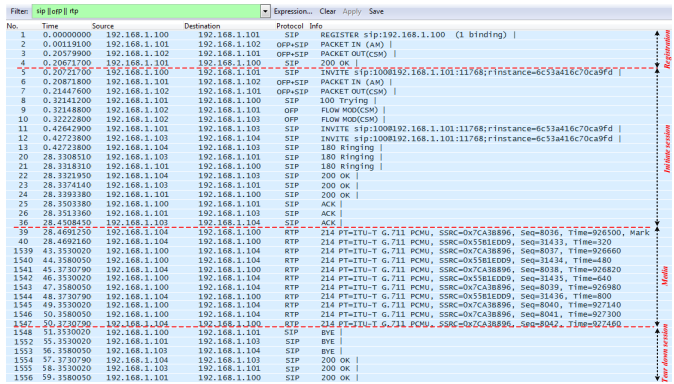

Fig. 29. Wireshark capture of advanced SIP call flow

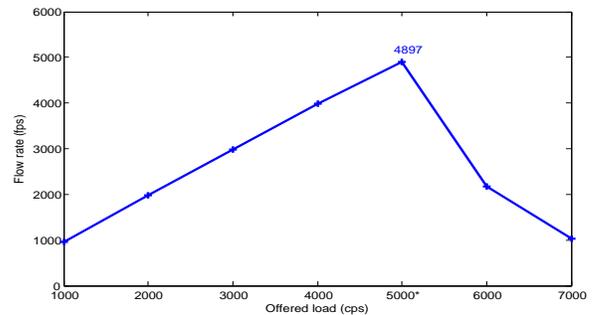

Fig. 30. OpenSIP$^{Full}$ controller throughput

TABLE V
OpenSIP$^{\text{Full}}$ controller response time

| Offered load (calls/sec) ~ | 1000 cps | 2000 cps | 3000 cps | 4000 cps | 5000* cps | 6000 cps | 7000 cps |
|---|---|---|---|---|---|---|---|
| Average response time (ms) | 9.37 | 17.36 | 29.47 | 52.46 | 86.45 | 693.45 | 947.53 |

TABLE VI
The characteristics of the OpenSIP$^{\text{NFV+}}$ implementation testbed

| | Physical Machines (PMs) | SDN, NFV Controller | OpenFlow Switches | Virtual Machines (VMs) | Traffic Generator (UAs) |
|---|---|---|---|---|---|
| Software | - | Floodlight v1.2 | Open vSwitch v2.4.1 | SIP Proxies (Kamailio v4.3.6) | SIPp v3.4 |
| Quantity | 4 | 1 | 16 | On demand | 1 |
| CPU | Intel Core i7-5960X 3GHz* | Intel Xeon E5645 2.4GHz | Intel Dual Core 1.8GHz | 1 vCPU ~ 1 Core | Intel Dual Core 1.8GHz |
| RAM | 32 GB | 4 GB | 1 GB | 2 GB | 2 GB |
| Operating System | CentOS v7.2 | Linux kernel v3.10 | Linux kernel v3.10 | CentOS v7.2 on Oracle's VirtualBox | Red Hat v6 |

\* 8 Cores

TABLE VII
OpenSIP$^{\text{NFV+}}$ throughput and the number of VMs required

| Time (sec) | 0 - 100 | 100 - 200 | 200 - 300 | 300 - 400 | 400 - 500 | 500 - 600 | 600 - 700 | 700 - 800 | 800 - 900 |
|---|---|---|---|---|---|---|---|---|---|
| Offered load (calls/sec) ~ | 1000 | 2000 | 3000 | 4000 | 5000 | 6000 | 4000 | 2000 | 1000 |
| Throughput (calls/sec) ~ | 989 | 1976 | 2898 | 3879 | 4875 | 5867 | 3978 | 1956 | 945 |
| The number of VMs | 1 | 2 | 3 | 4 | 5 | 6 | 4 | 2 | 1 |

TABLE VIII
Comparison between the different approaches with offered load (cps) ~ 3000

| Approaches | TLWL-1.75 | OpenSIP$^{\text{Partial}}$ | OpenSIP$^{\text{Full}}$ | OpenSIP$^{\text{NFV+}}$ |
|---|---|---|---|---|
| Throughput (calls/sec) ~ | 1785 | 2798 | 2841 | 2898 |
| Average end-to-end call setup time (ms) ~ | 38.74 | 29.98 | 29.47 | 28.31 |

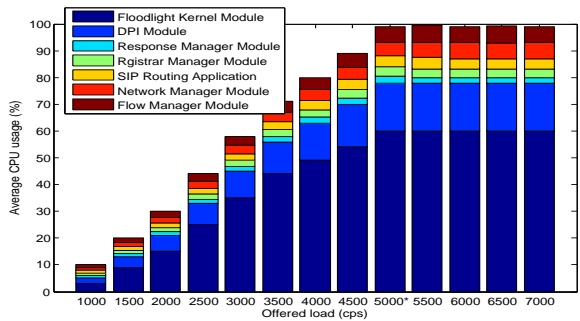

Fig. 31. OpenSIP$^{\text{Full}}$ controller CPU profile

algorithm proposed by Ohta [21] focuses on queue length. The drawback of local algorithms is the cost of call rejection. The proxy must use its resources for rejecting excess calls when dealing with heavy overloads.

In distributed algorithms, on the other hand, proxies collaborate together to control the overload, so that upstream proxies control the load of the downstream proxies (see [9], [22], [23]). Distributed algorithms are categorized into implicit and explicit methods. The absence of responses or the loss of packets is used to detect overload in implicit methods. Explicit algorithms are classified into rate-based, loss-based, signal-based, window-based, and on/off control methods. Among the rate-based methods, the downstream proxy controls the delivery rate of upstream proxies [22]–[24]. In loss-based method, the downstream proxy measures its current load and accordingly requests the upstream proxies to reduce their load. In window-based methods, unless there is an empty space in the upstream proxy window, the load is not transmitted to the downstream proxy. Window size adaptation can be achieved using feedback from the downstream proxy [9]. In signal-based algorithms, the upstream proxy reduces its transmission rate when receiving the `503 Service Unavailable` message [25]. By transmission of the *Retry-After* feedback [26], a proxy can either hold off or on to its received load within the on/off control method.

Based on the mentioned points, the disadvantages of the present approach to SIP overload control are complexity and overhead which results in instability.

Note that, NFV based approaches also provide a simple alternative to application-layer overload control.

Another approach for dealing with the overload issue is load balancing. This is the distribution of traffic among SIP proxies according to their available capacity by the use of a load balancer [17], [18]. Also, the DNS-based load balancing methods can provide an alternative to in-line load balancing, by associating multiple servers to the same SIP URI [27].

## VI. Conclusion and Future Work

SIP proxy overload is one of the main problems of SIP networks, which results in a sharp drop in performance due to both improper request routing in the application layer and resource saturation. In the current paper, this problem is overcome by three frameworks for upgrading SIP networks, which are based on SDN and NFV technologies (OpenSIP). One of the key challenges in SDN is the lack of application awareness in controllers or switches, which prevents them from making smart decisions. We used the DPI engine to achieve awareness of SIP message details. For this purpose, the OpenFlow protocol was also extended. In the OpenSIP$^{\text{Partial}}$, the call requests are distributed among SIP proxies by extending the OpenFlow controller and switch so that a minimum number of messages are exchanged between the switches and controller. In this method, the convenient proxy and the shortest network path are selected together. In the OpenSIP$^{Full}$, the control plane of the SIP proxy is decoupled from the infrastructure layer and the SIP routing operation is implemented in software in the form of a logically centralized controller. In the OpenSIP$^{NFV+}$, using the concept of NFV, the SIP proxy is virtualised to prevent hardware limitations. All three approaches were implemented by Floodlight and Open vSwitch tools and thorough performance evaluations were conducted through various scenarios. The simulation results confirmed that all OpenSIP implementations achieve a high throughput, low response time, and resource efficiency. The other achievements are the scalability of the proposed controllers and the proper routing of call requests.

One future work is to design a mathematical model for SIP routing in the controller, which can include constraints, such as link latency or network SLA. The VMs migrate technology can also be employed to manage virtual SIP proxies. Using the proposed approaches to provide security mechanisms for the SIP network can also be considered as a worthwhile future work. We also plan to extend OpenSIP to support encrypted traffics like SIP over TLS, as encryption is a major implementation requirement.

PLACE
PHOTO
HERE

**Ahmadreza Montazerolghaem** received the B.Sc. degree in Information Technology from the computer department, Sadjad University of Technology and M.Sc. degree in computer engineering from the computer department, Ferdowsi University of Mashhad (FUM), Iran, in 2010 and 2013, respectively. Currently, he is a Ph.D. candidate in computer engineering at computer department, FUM. He is an IEEE Student member and a member of IP-PBX type approval lab in FUM. He is also a member of National Elites Foundation (Society of prominent students of the country). His research interests are in Software Defined Networking, Network Function Virtualization, Voice over IP, and Optimization.

PLACE
PHOTO
HERE

**Mohammad Hossein Yaghmaee Moghaddam** received his B.S. degree in communication engineering from Sharif University of Technology, Tehran, Iran in 1993, and M.S. degree in communication engineering from Tehran Polytechnic (Amirkabir) University of Technology in 1995. He received his Ph.D degree in communication engineering from Tehran Polytechnic (Amirkabir) University of Technology in 2000. He has been a computer network engineer with several networking projects in Iran Telecommunication Research Center (ITRC) since 1992. November 1998 to July1999, he was with Network Technology Group (NTG), C and C Media research labs., NEC corporation, Tokyo, Japan, as visiting research scholar. September 2007 to August 2008, he was with the Lane Department of Computer Science and Electrical Engineering, West Virginia University, Morgantown, USA as the visiting associate professor. July 2015 to September 2016, he was with the electrical and computer engineering department of the University of Toronto (UoT) as the visiting professor. Currently, he is a full professor at the Computer Engineering Department, Ferdowsi University of Mashhad (FUM). His research interests are in Smart Grid, Computer and Communication Networks, Quality of Services (QoS), Software Defined Networking (SDN) and Network Function Virtualization (NFV). He is an IEEE Senior member and head of the IP-PBX type approval lab in the Ferdowsi University of Mashhad. He is the author of some books on Smart Grid, TCP/IP and Smart City in Persian language.

PLACE
PHOTO
HERE

**Alberto Leon-Garcia** received the B.S., M.S., and Ph.D. degrees in electrical engineering from the University of Southern California, Los Angeles, CA, USA, in 1973, 1974, and 1976, respectively. He was founder and CTO of AcceLight Networks, Ottawa, ON, Canada, from 1999 to 2002, which developed an all-optical fabric multiterabit, switch. He is currently a Professor in Electrical and Computer Engineering at the University of Toronto, ON, Canada. He holds a Canada Research Chair in Autonomic Service Architecture. He holds several patents and has published extensively in the areas of switch architecture and traffic management. His research team is currently developing a network testbed that will enable at-scale experimentation in new network protocols and distributed applications. He is recognized as an innovator in networking education. In 1986, led the development of the University of Toronto-Northern Telecom Network Engineering Program. He has also led in 1997 the development of the Master of Engineering in Telecommunications program, and the communications and networking options in the undergraduate computer engineering program. He is the author of the leading textbooks Probability and Random Processes for Electrical Engineering and Communication Networks: Fundamental Concepts and Key Architecture. His current research interests include application- oriented networking and autonomic resources management with a focus on enabling pervasive smart infrastructure. Prof. Leon-Garcia is a Fellow of the Engineering Institute of Canada. He received the 2006 Thomas Eadie Medal from the Royal Society of Canada and the 2010 IEEE Canada A. G. L. McNaughton Gold Medal for his contributions to the area of communications.